# A new artificial photosynthetic system coupling photovoltaic electrocatalysis with photothermal catalysis

Yaguang Li[1, 4, *], Fanqi Meng[2, 4], Xianhua Bai[1, 4], Dachao Yuan[3], Xingyuan San[1], Baolai Liang[1], Guangsheng Fu[1], Shufang Wang[1], Lin Gu[2], Qingbo Meng[2, *]

[1]Hebei Key Lab of Optic-electronic Information and Materials, The College of Physics Science and Technology, Institute of Life Science and Green Development, Hebei University, Baoding, 071002, China.
[2]Beijing National Laboratory for Condensed Matter Physics, Institute of Physics, Chinese Academy of Sciences, Beijing, 100190, China.
[3]College of Mechanical and Electrical Engineering, Hebei Agricultural University, Baoding 071001, China.
[4]These authors contributed equally to this work.

Correspondence and requests for materials should be addressed to Y. Li. (email: liyaguang@hbu.edu.cn) or to Q. Meng. (email: qbmeng@iphy.ac.cn).

**Abstract**

In this work, we present a novel artificial photosynthetic paradigm with square meter (m$^2$) level scalable production by integrating photovoltaic electrolytic water splitting device and solar heating CO$_2$ hydrogenation device, successfully achieving the synergy of 1 sun driven 19.4% solar to chemical energy efficiency (STC) for CO production (2.7 times higher than state of the art of large-sized artificial photosynthetic systems) with a low cost (equivalent to 1/7 of reported artificial photosynthetic systems). Furthermore, the outdoor artificial photosynthetic demonstration with 1.268 m$^2$ of scale exhibits the CO generation amount of 258.4 L per day, the STC of ~15.5% for CO production in winter, which could recover the cost within 833 suuny days of operation by selling CO.

**Introduction**

Artificial photosynthesis can convert CO$_2$ and H$_2$O into useful fuels, chemicals (CO,[1] CH$_4$,[2] etc.) and O$_2$ under solar irradiation, which is the most important way for carbon neutralization.[3-9] The application of large-scale artificial photosynthesis is of great significance to weaken the global warming, overcome the current energy and environmental crisis.[10-12] More recently, several large-sized artificial photosynthetic systems for CO$_2$ utilization have been reported, e. g., the solar fuel production chain with square meters (m$^2$) scale,[13] the photovoltaic electrocatalytic device with ~0.1 m$^2$ scale.[14] To the best of our knowledge, the highest solar to chemical energy efficiency

(STC) of large-sized devices is 7.2% by the photovoltaic electrocatalytic system.[14] However, the material cost for constructing large-sized artificial photosynthetic system is too expensive to practical application, due to the using of noble metal catalysts (e.g., Ir, Pt, Rh, Ru) and the costliness of large-sized components (e.g., membranes, solar reactor) in the devices.[15,16] Therefore, it is one of the holy grails of the entire scientific and technological community to achieve a scalable artificial photosynthetic system with high STC and low cost simultaneously, so as to realize the sustainable development of human society.

Herein, we have developed a new artificial photosynthetic system by integrating a photovoltaic electrolytic $H_2O$ decomposition part and a solar heating $CO_2$ hydrogenation part. Just relying on such a simple strategy, this system not only changed the reaction path and mass transportation but also discarded all rare elements and expensive components, resulting in the $m^2$ level scalable production and a record STC (19.4%) with low cost. Moreover, an outdoor demonstration (1.268 $m^2$ scale) of this new design was built based on full commercial components and the STC was still higher than 15% at outdoor test in winter. This artificial photosynthetic system could recover the total system cost within 833 days of operation by selling the products of CO.

**Conception for constructing novel artificial photosynthetic system**

It is well known that the widely studied photovoltaic electrocatalytic systems contain the competition of two main reactions: $H_2O$ decomposition and $CO_2$ hydrogenation on one system with $CO_2$ transportation through liquid electrolytes. Although various efficient catalysts have been developed, such as metals,[17-19] metal compounds,[20-22] molecular complexes,[23,24] photovoltaic electrocatalysis still faces two intrinsic shortcomings: one is the complex reaction processes in single catalytic system and the other is the sluggish $CO_2$ supply through gas/liquid transportation.[25-27] Here, a new paradigm of artificial photosynthesis is proposed to separate the two reactions of water splitting ($2H_2O \rightarrow 2H_2 + O_2$)[28] and $CO_2$ hydrogenation ($CO_2 + H_2 \rightarrow CO + H_2O$)[29-31] in space and time. There are four major advantages in this new system: (1) mature technologies can be selected for both water splitting and $CO_2$ hydrogenation; (2) the integrated system can be easily enlarged; (3) the systems for the two reactions can be optimized separately, providing a variety of possibilities for efficiency, cost and products; (4) $CO_2$ supply can be boosted by avoiding the gas transport in liquid elelctrolytes. As shown in Fig.1, this is an integrated system in which the hydrogen generated from photovoltaic water electrolysis[32] is directly injected into the solar heating system for $CO_2$ hydrogenation.[33-35] The $CO_2$ transportation of this system is in gas diffusion mode at a rate of $10^{-5}$ $m^2$ $s^{-1}$,[36] 10000 times higher than the rate of $CO_2$ diffused through liquid electrolytes ($10^{-9}$ $m^2$ $s^{-1}$)[37] in conventional photovoltaic electrocatalytic systems,[38,39] which could meet the $CO_2$ supply for large-sized artificial photosynthetic systems. For integrating such a new artificial photosynthetic system, the two issues should be solved firstly. One is the matching problem of solar energy utilization in this system, that is, how to scientifically distribute the proportion of solar energy irradiated to the two devices to improve the STC; the second is the quality matching of hydrogen production and hydrogen consumption in the new system.

**Integrating the artificial photosynthetic system**

A TiC/Cu heterostructure photothermal material was choose to construct the solar heating catalytic system (Supplementary Fig. 1),[40-42] which could heat the catalysts to 318 °C under 1 kW m$^{-2}$ intensity of sunlight (1 sun) irradiation to run $CO_2$ hydrogenation (Supplementary Fig. 2). This is the key for realizing the new artificial photosynthetic system, because the low solar irradiated temperature of conventional photothermal system (~80 °C, Supplementary Fig. 3) can not drive photothermal $CO_2$ hydrogenation under ambient solar irradiation. As the Fe single-atom catalysts (Fe SACs, Supplementary Fig. 4-8) were used as catalysts for solar heating $CO_2$ hydrogenation, the system showed a CO generation rate of 21.14 mol m$^{-2}$ h$^{-1}$ under 1 sun irradiation, corresponding to 24.1% of solar to chemical energy efficiency (detailed calculation seen in Supplementary Methods, Supplementary Fig. 9). More interestingly, as the 1% $O_2$-polluted $H_2$ was used as feed gas, the efficiency of $CO_2$ hydrogenation had little change (Supplementary Fig. 9a) and this solar heating system still showed a straight forward CO production rate of ~21.1 mol m$^{-2}$ h$^{-1}$ more than 60 days (Supplementary Fig. 10), evidencing the robustness of solar heating catalytic system.

The low requirement of hydrogen purity for solar heating catalysis enables us to simplify the photovoltaic electrocatalysis. Besides using commercialized single crystalline silicon solar cells (23.5% efficiency) as electric energy supply, the membrane was eliminated from the electrocatalytic reactor (Fig. 1) and the cheap nickel-plated stainless-steel mesh (Ni/stainless steel, Supplementary Fig. 11) was used as the electrodes to replace the precious electrocatalysts.[17-19] In the membrane free electrocatalytic reactor, the Ni/stainless steel's electrodes could achieve a current density of 10 mA cm$^{-2}$ in 1 M KOH electrolyte at only 1.53 V (Supplementary Fig. 12). The H$_2$ production rate of this photovoltaic electrolytic system was 2.40 mol m$^{-2}$ h$^{-1}$ under 1 sun irradiation (Supplementary Fig. 13), equivalent to 19.04% solar to hydrogen chemical efficiency (detailed calculation seen in Supplementary Methods). It was calculated that the solar cell's electric energy to chemicals energy efficiency (EC) of this photovoltaic electrocatalytic water splitting system was 81% (detailed calculation seen in Methods). The released H$_2$ contained ~0.8% O$_2$, which also meets the purity requirement of solar heating CO$_2$ hydrogenation.

Based on the above experimental resluls, the photovoltaic electrolytic water splitting device with 2800 cm$^2$ of solar irradiation area and solar heating CO$_2$ hydrogenation device with 240 cm$^2$ of solar irradiation area were integrated as a new type of artificial photosynthetic system with more than 3000 cm$^2$ of solar irradiation area in the laboratory (Fig. 1).

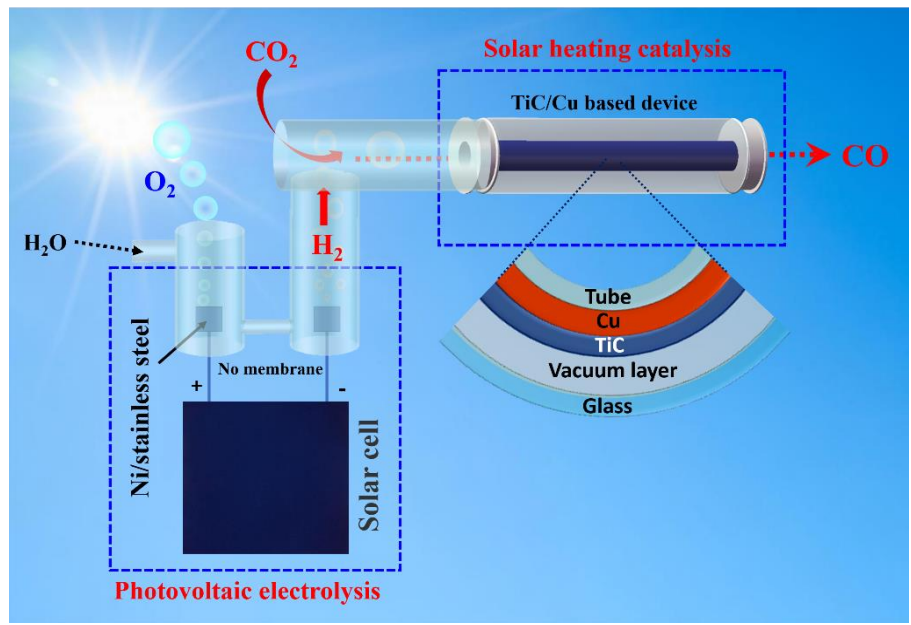

**Fig. 1** Schematic map of the novel artificial photosynthesis.

**The performance of novel artificial photosynthesis**

Fig. 2a showed that the laboratory system could produce CO with a rate of 38, 210, 491 mmol h$^{-1}$ under 0.6, 0.8, 1 sun irradiation, respectively. Additionally, Fig. 2b identified that this system showed a 100% selectivity for $CO_2$ converted into CO under different intensities of solar irradiation due to the +3 oxidation state of Fe-SACs (Supplementary Fig. 14).[43] Fig. 2c illustrated that the STC of new artificial photosynthetic system was increased from 11.3%, 17.4% to 19.4% along with the 0.6, 0.8 to 1 sun irradiation (Detailed calculation seen in Methods), which was 2.7 times higher than the best record value of scalable artificial photosynthesis with ~1000 cm$^2$ of solar irradiation area (7.2%).[14] The $CO_2$ reduction performance of this system was continuously tested for 6 days, and the CO production rate was stable maintained at ~500 mmol h$^{-1}$ (Fig. 2d), indicating the excellent stability of new artificial photosynthetic system.

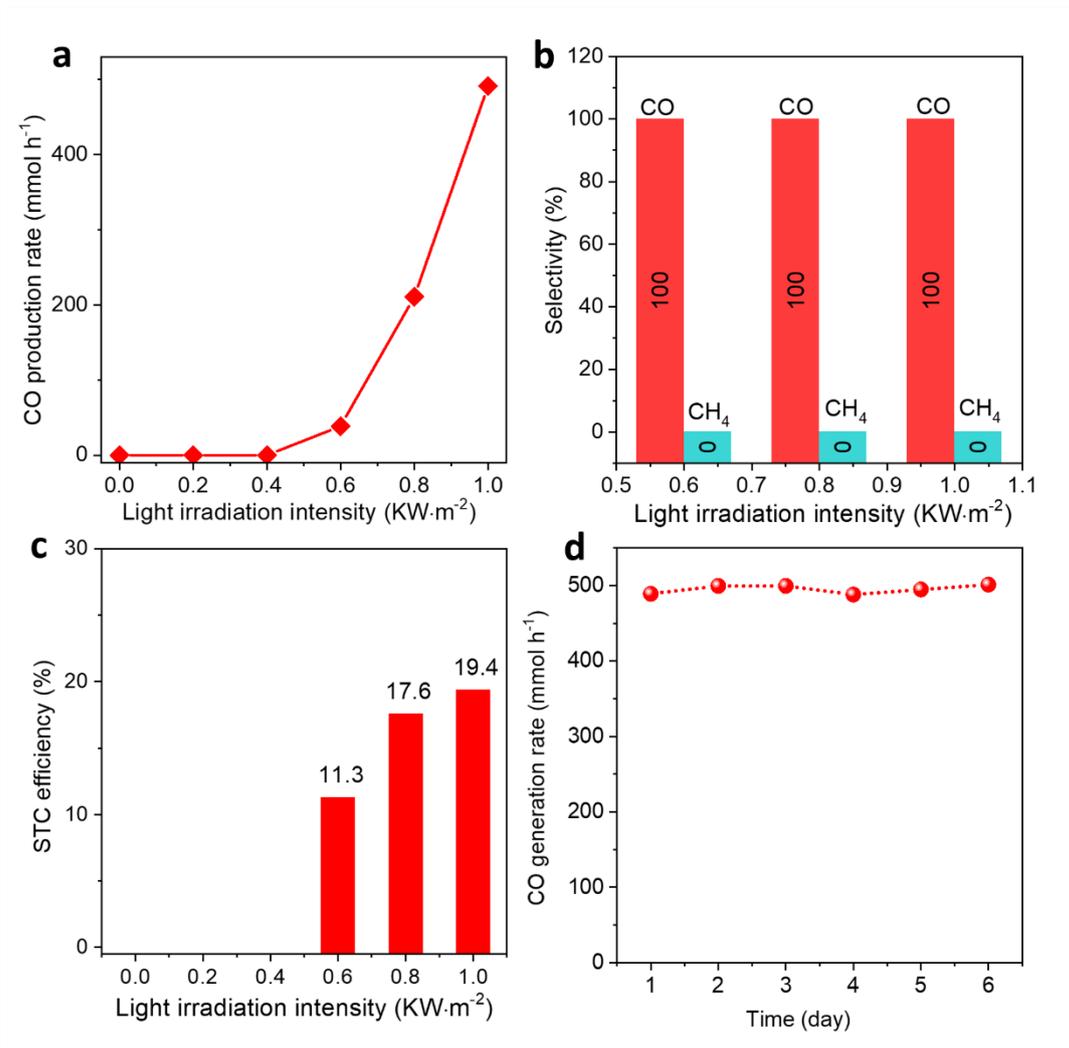

**Fig. 2** The laboratory performance of novel artificial photosynthetic system. **a** The CO production rate of new artificial photosynthetic system with Fe SACs, under different intensities of solar irradiation. **b** The CO selectivity of new artificial photosynthetic system with Fe SACs, under different intensities of solar irradiation. **c** The STC efficiency of new artificial photosynthetic system with Fe SACs, under different intensities of solar irradiation. **d** The CO production rate stability of new artificial photosynthetic system under 1 sun irradiation.

**The outdoor artificial photosynthetic demonstration**

The commercial singlecrystalline silicon solar cell panel (1.07 m² scale), membrane-free electrolytic water splitting device and factory prepared TiC/Cu based solar heating tube were used to build the outdoor artificial photosynthetic system. For maintaining the solar heating system at high temperature all day, a parabolic reflector with 0.198 m$^2$ of irradiated area (Supplementary Fig. 15) was applied to concentrate outdoor sunlight on solar heating device (Fig. 3a). A commercial $CuO_x/ZnO/Al_2O_3$ (SCST-401, Supplementary Fig. 16) was selected as the catalyst for solar heating reverse water-gas-shift reaction ($CO_2+H_2 \rightarrow CO+H_2O$). In outdoor test, the membrane-free electrolytic water splitting device was driven by the silicon solar cell panel to produce $H_2$, then the $H_2$ and $CO_2$ entered into the solar heating system for $CO_2$ hydrogenation. The artificial photosynthetic system for CO production was tested in December 20, 2021, with an ambient temperature of 2~13 °C and a solar irradiation intensity of 0.26-0.49 kW m$^{-2}$ in the daytime in Baoding City of Hebei Province, China. As shown in Fig. 3c, the CO generation occurred at 9:00 AM with a production rate of 27.9 L h$^{-1}$. After that, the CO generation rate rose to a peak value of 41.4 L h$^{-1}$ at 12:00 PM and then gradually decreased to 23.6 L h$^{-1}$ at 16:00 PM. The total amount of CO produced daily was up to 258.4 L. Although the solar intensity and ambient temperature are the lowest in winter, the outdoor system STC for CO production was still in the range of 15% to 15.8% throughout the operating period (Fig. 3d, detailed calculation seen in Methods).

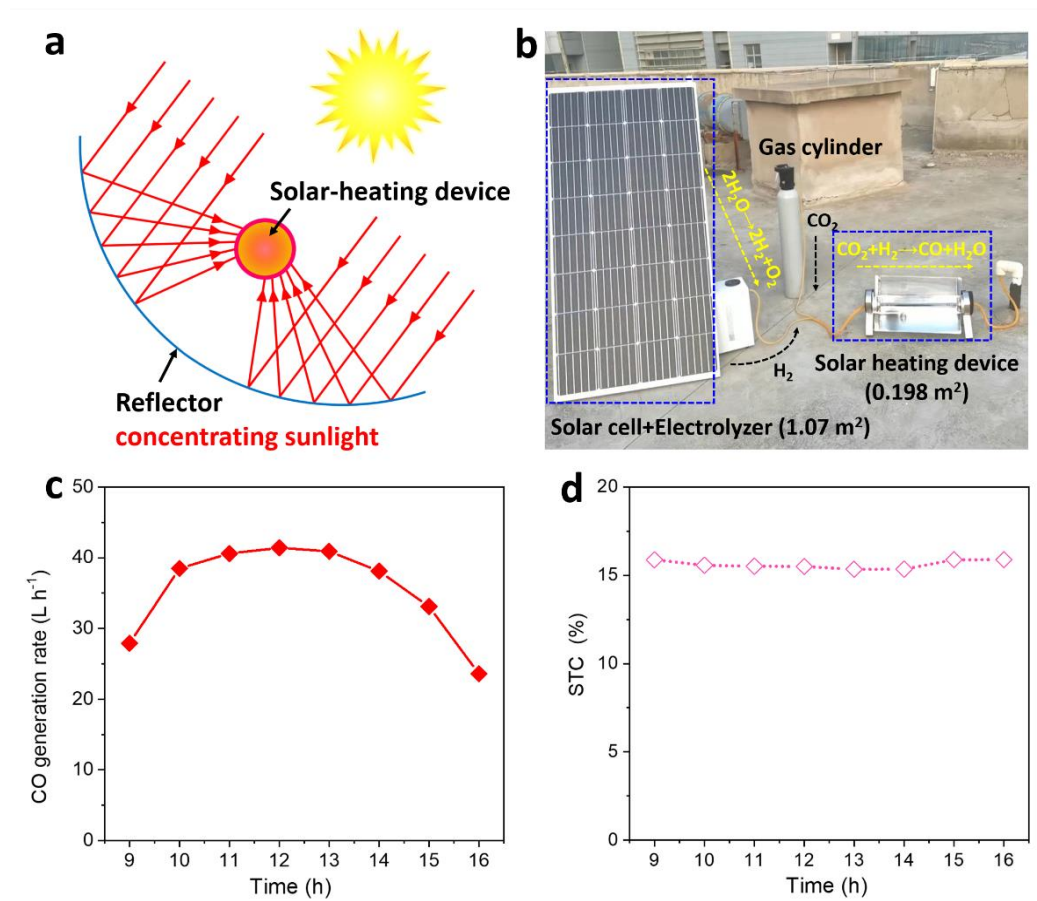

**Fig. 3** The outdoor performance of novel artificial photosynthetic system. **a** The location diagram of reflector, solar heating device and solar. **b** The photograph of new artificial photosynthetic demonstration on the roof of the building in Hebei University. **c, d** The CO production rate and STC of new artificial photosynthetic demonstration under ambient sunlight irradiation, on December 20, 2021, in Baoding City, China.

Tab.1 listed the data of new artificial photosynthetic systems and the most advanced large-sized artificial photosynthetic systems. Firstly, the size of the outdoor artificial photosynthetic system was 1.268 m², and all parts can be processed in the factory, showing that the system could realize mass production directly. Secondly, the STCs of lab and outdoor systems for $CO_2$ reduction as CO were 19.4% and 15-15.8% respectively, which was 2.7 times and 4 times higher than that of reported large-sized artificial photosynthetic systems under lab and outdoor conditions, respectively.[13,14] The total cost of outdoor demonstration was calculated as $1018 per $m^2$ (Supplementary Fig. 17). Tab. 1 showed that the cost of large-sized artificial photosynthetic devices is too expensive to calculated cost.[13,14,44] As far as we known, the cheapest cost of small-sized artificial photosynthetic device reported in literatures was ~$7200 per $m^2$ (Tab. 1),[45] which was 7 times higher than our outdoor demonstration. With the ultra-high STC and ultra-low system cost, the system cost recovery time of the outdoor artificial photosynthetic device was calculated by selling product (CO). Referring to the price of CO ($6 per $m^3$),[45] the outdoor system could recover the cost after 833 days of operation, which corresponds to ~3.5 years (detailed calculation seen in Methods). The service life of the components in this outdoor system for CO production was generally more than 10 years, able to profitable by selling CO.

**Tab. 1** Comparison of the solar driven $CO_2$ reduction systems of this work and the state of the art of solar cells driven artificial photosynthetic systems. IA is the working solar illumination area of device used for $CO_2$ conversion.

| Refs. | Condition | IA ($cm^2$) | Main product | STC (%) | Cost ($ per $m^2$) |
|---|---|---|---|---|---|
| This work | outdoor | 12,680 | CO | 15-15.8 | 1018 |
| 13 | outdoor | 65,000 | CO+$H_2$ | 3.8 | None |
| 44 | outdoor | 14,700 | formic acid | 1.86 | None |
| This work | Lab | 2,287 | CO | 19.4 | None |
| 45 | Lab | 14 | CO | 8.05 | 7200 |
| 14 | Lab | 987 | formic acid | 7.2 | None |
| 46 | Lab | 16 | CO+$H_2$ | 4.3 | None |

## Conclusion

In this work, a novel artificial photosynthesis paradigm was proposed, in which the silicon solar cells were used to drive the membrane free electrolyzer for photovoltaic electrolytic water splitting as $O_2$ and $H_2$. Then, the generated $H_2$ and $CO_2$ were injected into the solar heating system based on a TiC/Cu based device to carry out efficient sunlight driven $CO_2$ hydrogenation due to the high 1 sun-heating temperature of 318 °C. The photovoltaic electrolytic reactor eliminated the membrane and used the Ni-plated stainless steel mesh as the electrodes to reduce the cost. As the 240 cm$^2$ of solar heating $CO_2$ hydrogenation device was integrated to 2800 cm$^2$ of silicon solar cell driven photovoltaic electrolytic water splitting device, the system exhibited a $CO_2$ conversion rate of 491 mmol h$^{-1}$, an STC of 19.4%, a selectivity of 100% for CO production, under 1 sun irradiation. Moreover, an outdoor demonstration with 1.268 m$^2$ of solar irradiation area was constructed, which showed a cost of \$1018 per m$^2$, the gas production of 258.4 L per day, the STC of 15%-15.8% for CO production in winter, under ambient solar irradiation, which could neutralize device cost by selling the product of CO within 833 sunny operation days, revealing the ability for direct scalable application.

## Outlook

The new artificial photosynthetic system has huge space for STC improvement and flexible product regulated ability. As the silicon solar cell was replaced by triple-junction solar cells for photovoltaic electrocatalytic water splitting, the calculated STC of new artificial photosynthetic system was as high as 28.9 % (detailed calculation seen in Supplementary Methods), which was higher than the best STC (19.1%) of triple-junction solar cells driven artificial photosynthesis.[47] Further, this system could convert product from CO to $CH_4$ by changing the solar heating $CO_2$ hydrogenation catalysts as commercial $Ni/Al_2O_3$ (Supplementary Fig. 18, detailed calculation seen in Methods). Therefore, this system could be a core platform for scientists all over the world to realize carbon neutralization via converting $CO_2$ and $H_2O$ into a variety of chemicals.

## Methods

**Thermocatalytic $CO_2$ hydrogenation**

The thermocatalytic activity of catalysts for $CO_2$ hydrogenation was tested by the fixed-bed reactor (XM190708-007, DALIAN ZHONGJIARUILIN LIQUID TECHNOLOGY CO., LTD) in continuous flow form. Typically, 10 mg of catalyst was placed in a quartz flow reactor. For CO production, the feed gas of $CO_2/H_2/Ar = 1/1/48$ or $CO_2/99\% H_2+1\% O_2/Ar = 1/1/48$ with 100 Sccm of flow rate was regulated by the mass flow controller. The reaction products were tested by gas chromatograph (GC) 7890A equipped with FID and TCD detectors.

**Solar heating $CO_2$ hydrogenation as CO**

The solar heating $CO_2$ hydrogenation was tested as follows: 137 g of Fe SACs were loaded into TiC/Cu based device (0.024 $m^2$), and irradiated by a xenon lamp (ZSL-4000). In this test, $CO_2$ and 100% $H_2$ (or 99% $H_2$+1% $O_2$) with 1.5/1 ratio were mixed as feed gas. For the produced gas, the flow rate was tested by mass flowmeter and the composition was tested by GC 7890A equipped with FID and TCD detectors. It was required to control the flow rate to make the $H_2$ consumption exceeds 95%. The data were collected by FID and TCD.

The CO rate ($\delta$, mol m$^{-2}$ h$^{-1}$) was calculated as follows:

$$\delta \text{ (mol m}^{-2}\text{ h}^{-1}) = (L/(24.5*S)) \qquad (1)$$

L was the CO flow rate (L h$^{-1}$). S was the irradiated area (0.024 m$^{-2}$). When using the 99% $H_2$+1% $O_2$ as feed gases, the L irradiated by 0.6, 0.8, 1 sun was 0.98, 5.32, 12.43 L h$^{-1}$, respectively.

**Solar driven water splitting**

The back contact silicon cells interdigitated with 2800 cm$^2$ irradiation area were purchased from SUNPOWER (23.5% efficiency) to drive an alkaline electrolyzer with 1 m$^2$ of Ni/Stainless steel mesh. Xenon lamp (HP-2-4000) was used as a light source and 1M KOH was used as the electrolyte for sunlight driven water splitting. The produced $H_2$ was injected into the TiC/Cu based device and the produced $H_2$ rate of the solar driven water splitting system was tested by mass flowmeter (C50 300SCCM).

The $H_2$ production rate for per m$^2$ (H, mol m$^{-2}$ h$^{-1}$) of solar cell was calculated as follows:

$$H \text{ (mol m}^{-2}\text{ h}^{-1}) = (\varepsilon/(24.5*S)) \qquad (2)$$

$\varepsilon$ (L h$^{-1}$) was the $H_2$ generation amount per hour detected by a flowmeter, S was the irradiated area (0.2800 m$^{-2}$). The $\varepsilon$ irradiated by 0.4, 0.6, 0.8, 1 sun was 6.68, 10.02, 13.31, 16.45 L h$^{-1}$, respectively.

**Enthalpy change energy of chemicals**

The enthalpy change energy of $CO_2$ (g), CO (g), $H_2$ (g), $O_2$ (g), $H_2O$ (g), $H_2O$ (l) was -

393.505, -110.541, 0, 0, -241.818, -285.830 kJ mol$^{-1}$, respectively.

The (g) and (l) indicated the gas state and liquid state, respectively.

**Novel artificial photosynthesis for $CO_2$ and $H_2O$ converted as CO and $O_2$**

As the solar driven water splitting produced $H_2$ injected into the TiC/Cu based device loaded with 137 g of Fe SACs, 600 sccm of $CO_2$ was simultaneously put into the TiC/Cu based device, which was controlled by mass flow controller (C50 5SLM). The TiC/Cu based device was irradiated by a xenon lamp (ZSL-4000). For the produced gas, the flow rate was tested by mass flowmeter (C50 5SLM) and the composition was tested by GC 7890A equipped with FID and TCD detectors.

The CO rate ($\delta$, mmol h$^{-1}$) was calculated as follows:

$$\delta \text{ (mmol h}^{-1}) = (1000*L/24.5) \qquad (3)$$

L was the CO flow rate (L h$^{-1}$) and the L irradiated by 0.6, 0.8, 1 sun was 0.946, 5.170, 12.030 L h$^{-1}$, respectively.

**The STC calculation of sunlight driven $CO_2$ conversion as CO**

The STC efficiency of novel artificial photosynthetic system for converting $CO_2$ into CO was calculated as follows:

$$STC = (\Delta H * \varepsilon)/(I*S*3600) \qquad (4)$$

$\Delta H$ was the reaction Enthalpy change energy ($H_2O$ (l) + $CO_2$ (g) → CO (g) + 1/2 $O_2$ (g) + $H_2O$ (g), $\Delta H$= 326.9754 kJ mol$^{-1}$), $\varepsilon$ (mol) was the CO generation amount per hour detected by a flowmeter, I was the light intensity (kW m$^{-2}$), S was the total irradiated area. The $\varepsilon$ irradiated by 0.6, 0.8, 1 sun was 0.0386, 0.211, 0.491 mol, respectively.

Since not all $H_2$ produced from solar driven water splitting was used for $CO_2$

hydrogenation, the irradiation area (β) of solar driven water splitting used for $CO_2$ hydrogenation was calculated as follows:

β = M/N*0. 2800 m$^{-2}$    (5)

The M was $H_2$ used for $CO_2$ hydrogenation as CO, which was equal to the CO production rate of 0.0386, 0.211, 0.491 mol h$^{-1}$, under 0.6, 0.8, 1 sun irradiation, respectively. The N was the $H_2$ production rate of 0.3931, 0.5273, 0.6714 mol h$^{-1}$, irradiated by 0.6, 0.8, 1 sun, respectively. Therefore, the β was 0.0276 m$^{-2}$, 0.1119 m$^{-2}$, 0.2047 m$^{-2}$, under 0.6, 0.8, 1 sun irradiation, respectively.

And the S = β +0.0240 m$^{-2}$    (6)

Therefore, the S was 0.0516, 0.1359, 0.2287 m$^{-2}$, under 0.6, 0.8, 1 sun irradiation, respectively.

Consequently, the STC was 11.3%, 17.6%, 19.4%, under 0.6, 0.8, 1 sun irradiation, respectively.

**The EC calculation**

The 1 sun driven EC of photovoltaic electrocatalytic water splitting in this work and reported photovoltaic electrocatalytic $CO_2$ reduction was calculated as follows:

EC= STCE/efficiency    (7)

The STCE was the solar to hydrogen chemical efficiency (19.04%) under 1 sun irradiation. The efficiency was the electric energy generation efficiency of solar cell (23.5%) under 1 sun irradiation. Therefore, the EC was calculated as 81.0%.

**Outdoor artificial photosynthetic system**

The outdoor artificial photosynthetic system consisted of two components. One component was photovoltaic electrolysis system, in which the PERC solar cell (182DCB) with 1.07 m² of solar irradiation area was used to power electrolytic reactor with 2.782 m² of Ni/Stainless steel mesh divided into 12 independent chambers in series. The mixture of 200 g KOH and 1.8 L deionized water was used as the electrolyte. The other component was solar heating system, in which a solar heating device was provided by Hebei scientist research experimental and equipment trade Co., Ltd. with the size of 4 cm in diameter and 50 cm inlength, eqquipped with a reflector of 50 cm inlength and 36 cm in width. For the production of CO, the catalysts used in solar heater were 400 g $CuO_x/ZnO/Al_2O_3$. For CO production production in solar heating system, the $CO_2/H_2$ ratio was 1.5. It was required to control the flow rate to ensure the $H_2$ consumption exceeds 95%. The data were collected by FID and TCD. The data shown in Fig. 3c were tested on December 20, 2021, in Baoding, China.

**The STC of outdoor artificial photosynthetic system**

The STC of the outdoor artificial photosynthetic system for converting $CO_2$ into CO was calculated as follows:

$$STC = (\Delta H * \varepsilon)/(I * S * 3600 * 22.4) \qquad (8)$$

$\Delta H$ was the reaction Enthalpy change energy ($H_2O$ (l) + $CO_2$ (g) → CO (g) + 1/2 $O_2$ (g) + $H_2O$ (g), $\Delta H$ = 326.9754 kJ mol$^{-1}$), $\varepsilon$ (L) was the CO generation amount per hour detected by a flowmeter, I was the outdoor solar intensity (kW m$^{-2}$), S was the total irradiated area of 1.268 m².

**The cost recovery calculation**

We assumed that the CO production amount of the outdoor artificial photosynthetic system was 258.4 L/day. Due to the variety of CO prices, the quotation of Chae et al. reported result[45] and North Special Gas Co., Ltd. was adopted, which was \$6 per m$^3$ CO.

Therefore, the income of outdoor artificial photosynthetic system for CO production was 0.2584* \$6=\$1.55. To achieve an income of \$1291, this system required \$1291/\$1.55=833 sunny days, which were equivalent to the sunny days in 3.5 years, according to the weather in Baoding of 240 sunny days per year.

## Data availability

The data that support the findings of this study are available from the corresponding authors upon reasonable requests.

## Acknowledgements

This work was supported by the Hebei Natural Science Foundation (Grant No. B2021201074), the Hebei Provincial Department of Science and Technology (Grant No. 216Z4303G), Hebei Education Department (Grant No. BJ2019016), the Advanced Talents Incubation Program of Hebei University (Grant Nos. 521000981248 and 8012605), the National Nature Science Foundation of China (Grant Nos. 51421002, 51702078, 61774053, 61504036, 51972094, and 51971157), the Natural Science Foundation of Hebei Province (Grant Nos. B2021201034, F2019201446, and F2018201058), the National Key Research and Development Program of China (2018YFB1500503-02), the Scientific Research Foundation of Hebei Agricultural University (YJ201939). Thank you for the TEM technical support provided by the Microanalysis Center, College of Physics Science and Technology, Hebei University.

## Author contributions

Y. Li, L. Gu, S. wang and Q. Meng conceived the project and contributed to the design of the experiments and analysis of the data. Q. Meng proposed the system coupling idea; Y. Li provided the solar heating system strategy. X. Bai and D. Yuan performed the TiC/Cu based device preparation and characterizations. X. Bai, and F. Meng performed the catalyst preparation and characterizations. B. Liang and G. Fu provided the optical advice. F. Meng and X. San conducted the SEM and TEM examinations. Y. Li and Q. Meng wrote the paper. All the authors discussed the results and commented on the manuscript.

## Additional information

**Supplementary Information** accompanies this paper

**Competing interests:** The authors declare no competing financial interests.

**Reprints and permission** information is available online